\documentclass[Review, times]{mysagej}

\usepackage{moreverb,url}

\usepackage{selinput}

\usepackage[colorlinks,bookmarksopen,bookmarksnumbered,citecolor=red,urlcolor=red]{hyperref}

\newcommand\BibTeX{{\rmfamily B\kern-.05em \textsc{i\kern-.025em b}\kern-.08em
T\kern-.1667em\lower.7ex\hbox{E}\kern-.125emX}}

\def\volumeyear{2020}

\SelectInputMappings{%
  aacute={á},
  ntilde={ñ},
}

\begin{document}

\runninghead{Ferrari et al.}

\title{Modeling provincial Covid-19 epidemic data in Italy using an adjusted time-dependent SIRD model}

\author{Luisa Ferrari\affilnum{1}, Giuseppe Gerardi\affilnum{1}, Giancarlo Manzi\affilnum{1}, Alessandra Micheletti\affilnum{2}, Federica Nicolussi\affilnum{1}, Elia Biganzoli\affilnum{3} and Silvia Salini\affilnum{1}}

\affiliation{\affilnum{1} Department of Economics, Management and Quantitative Methods, University of Milan, Milan, Italy\\
\affilnum{2} Department of Environmental Science and Policy,  University of Milan, Milan, Italy\\
\affilnum{3} Department of Clinical Sciences and Community Health, University of Milan, Milan, Italy}

\corrauth{Giancarlo Manzi, Department of Economics, Management and Quantitative Methods,
University of Milan,
Via Conservatorio, 7,
20122, Milan, Italy.}

\email{giancarlo.manzi@unimi.it}

\begin{abstract}
In this paper we develop a predictive model for the spread of COVID-19 infection at a provincial (i.e. EU NUTS-3) level in Italy by using official data from the Italian Ministry of Health integrated with data extracted from daily official press conferences of regional authorities and from local newspaper websites. This integration is mainly concerned with COVID-19 cause specific death data which are not available at NUTS-3 level from open official data data channels. An adjusted time-dependent SIRD model is used to predict the behavior of the epidemic, specifically the number of susceptible, infected, deceased and recovered people. Predictive model performance is evaluated using comparison with real data.
\end{abstract}


\maketitle

\section{Introduction}
The outbreak of the Covid-19 epidemics in early 2020 has caused an unprecedented effort of the scientific community to produce models that could monitor and predict the evolution of the epidemics in a reliable way, also to advice governments to take actions which could mitigate the burden of hospitals to treat the affected patients, and reducing the mortality rate of the infection. 

The first reported Italian case of Covid-19 dates back to February 20th, 2020 \cite{Guzzettaetal}, in the city of Codogno, southern Lombardy, and the epidemics spread particularly in Italian northern regions, that is, those most commercially connected with China, where the epidemics had its origin. The Italian government took subsequent measures to contain the epidemics \cite{Gattoetal}, ending soon with a full national lockdown on March 11th, 2020, to drastically reduce the mobility of citizens and the consequent infectious contacts. 

We decided to focus on the modelling of the epidemics in Italian provinces (i.e. at EU NUTS-3 level), rather than in Italian regions (i.e. at EU NUTS-2 level). This choice was dictated by the fact that the Covid-19 outbreak in Italy has been not homogeneously spread within regions, with many differences from province to province in the same region.

According to the characteristics of the virus spreading, it is not suited thinking of a uniform virus propagation behavior at the regional level. Even the timing of the initial stages of the infection and its dynamics seem to have been very different even among contiguous provinces, as clusters of Covid-19 contagion have been often located in very restricted areas. One of the proof of this lack of spread homogeneity is in the initial stage of the virus outbreak which was located at the border between Lombardy and Emilia-Romagna regions, namely in the Lodi province (Lombardy) and in Piacenza province (Emilia-Romagna). The provinces of Varese in Lombardy and Ravenna in Emilia-Romagna are not in the least comparable with the provinces of Lodi and Piacenza in terms of intensity of the contagion. Moreover, viral RNA swab tests had been initially conducted depending on the choices of the local health authorities, and hospital admissions in the early stages of the emergency depended on the management and absorption capacity of the local health units, resulting in many differences in practicing hospital care at a local level. 
Also the management of elder people, in particular those living in retirement homes, was quite different locally, causing in some cases the development of local surge of the infection and increase in mortality, since the illness was particularly severe on elder people.

The spread of epidemic is often effectively modeled by compartmental deterministic models \cite{Capasso,Diekmannetal}, where a population of susceptible individuals evolves into other categories representing the different stages of the infection. We consider here a model consisting of 4 compartments:  susceptible (S), infected (I),  recovered (R) and died (D), that were the only compartments for which we could find available data at NUTS-3 level in Italy. 
In the case of SARS-Cov-2 virus, it was proven that the infection has an incubation period of about 5 days and that a significant percentage of the infected people are asymptomatic \cite{Lietal,Vo, Huangetal}, thus actually more compartments should and have been considered both in deterministic and stochastic models (see e.g. \cite{Arenasetal,Bertuzzoetal,Gaeta,Gattoetal,Giordanoetal,ImperialCollege,Sebastianietal}). Unfortunately the data unavailability at NUTS-3 level would cause problems in the parameters identification \cite{Roosa}. Furthermore we decided to keep the model as simple as possible in order to make it more accountable and, at the same time, robust to the variation in time of the parameters.

The parameter of epidemics are actually evolving in time. In fact, the rate of infection depends implicitly on the limitation of the mobility of the citizens; on the measures of protection of the healthcare personnel and of the workers who kept on doing jobs which were considered essential services to the community; on the number of swab tests performed locally to detect the infected subjects in order to put them in strict quarantine. Also the recovery rate and the death rate are changing in time and in space because of the different burden of the local healthcare systems and the new insights in the pulmonary illness caused by the SARS-Cov-2 virus and its possible pharmacological and medical treatment. We thus studied a SIRD model whose parameters are evolving in time, and able to automatically adapt to the factors which are implicitly causing their changes. This makes our model able to predict with a good reliability the short term evolution of the epidemics, in particular in absence of sudden big changes of the population behavior or the health policy. We leave the detection of major change points in the parameters behavior to subsequent papers, using time series or stochastic processes techniques like the one described in \cite{Michelettietal}

In the following sections we describe our main data sources, the model that we implemented and its parameters estimation and provide a detailed study of the fitting of the predictions made by our model with real data.

The rest of the paper is organized as follows. Section 2 describes the data extraction we performed, including the extra-data on deaths not available at a province level from official authorities, together with a short discussion on issues in the data collection. Section 3 presents the model we used and some necessary adjustments on it in order to include time-depending events such as central government and local authorities lockdown decisions all along the outbreak. Section 4 describes an adjusted training process for the model we adopted to ameliorate the estimates. Section 5 is about the estimation performance of the model and Section 6 concludes the paper and outlines some possible future work.

\section{Data on Covid-19 in Italy}
Only data on Covid-19 daily cases were made available for each province (i.e. NUTS-3) by the Italian Presidency of the Council of Ministers (PCM) - Department of Civil Protection Agency (CPA), the Italian Ministry of Health and the Italian National Institute of Health, which are the official data providers for the Covid-19 outbreak in Italy. No data about deaths and recovered patients were provided\footnote{Official data are made available by CPA at \url{https://github.com/pcm-dpc/COVID-19}.}. Therefore, we decided to integrate the official data on cases with data on deaths derived from press conferences and reports published online by regional authorities or local newspapers. The number of the cumulative recovered individuals at a provincial level has been estimated using the recovery rate at the regional level, computed as the ratio of recovered people and the total number of cases in each day. Data on deaths have been scraped using the daily press conferences and Covid-19 bulletins for a vast majority of the Italian provinces.

Regions for which we were not able to obtain provincial death data from official bulletins or press conferences were Liguria, Lombardia, Veneto, Friuli-Venezia-Giulia and Campania. However, for Imperia province in Liguria region and Cremona province in Lombardy we were able to obtain data on Covid-19 deaths from local newspapers. Table \ref{tab1} contains all the sources we scraped for obtaining provincial Covid-19 death data. For Veneto region it was possible to obtain data on Veneto provincial Covid-19 deaths but only on request and with a considerable delay.

\begin{table}[h]
 \small\sf\centering
  \caption{Main data sources for provincial Covid-19 deaths}
    \begin{tabular}{|l|l|}
    \toprule
          Region & Main source \\
          \midrule
          Valle d'Aosta & https://github.com/pcm-dpc/COVID-19  \\
          Piemonte & https://www.regione.piemonte.it\\
          Lombardia & https://laprovinciacr.it \\
          & (data for Cremona province only)\\
          Veneto & no available data\\
          Friuli-Venezia-Giulia & no available data\\
          Trentino-Alto-Adige & https://github.com/pcm-dpc/COVID-19 \\
          Emilia-Romagna & https://www.regione.emilia-romagna.it \\
          Liguria & http://sanremonews.it  \\
          & (data for Imperia province only) \\
          Toscana & https://www.toscana-notizie.it/ \\
          Marche & http://www.regione.marche.it/\\
          Umbria & http://www.regione.umbria.it\\
          Lazio & https://www.facebook.com/SaluteLazio \\
          Abruzzo & https://www.regione.abruzzo.it\\
          Molise & http://www3.regione.molise.it\\
          Campania & no available data\\
          Puglia & http://www.regione.puglia.it \\
          Basilicata & https://www.regione.basilicata.it \\
          Calabria & https://portale.regione.calabria.it \\
          Sicilia & http://pti.regione.sicilia.it \\
          Sardegna & https://www.regione.sardegna.it \\
     \bottomrule
     \end{tabular}
     \label{tab1}
 \end{table}

Another series not available at a provincial level was that of the number of recovered people at time $t$. 
The reason for the choice of estimating this number proportionally to that of the region is that patient treatment for the illness due to Covid-19 could be considered more uniform across the provinces (with almost the same recovery rate across provinces within the region) than for the number of deaths.

Official data on the Covid-19 outbreak in Italy present many issues mainly related to delays in reporting new cases and deaths, incongruities (for example, negative values in the series of new cases due to post-event recounts), and missing data. Menchetti and Noirjean \cite{Menc} reported widely on the flaws and biases of these official data. Bartoszek et al. \cite{Bartoszek} highlighted that reporting statistics at a certain spatial level (national, regional, etc.) in Italy does not say much about the dynamics of the disease at lower levels. The problem of unreliable data becomes even more cogent with epidemiological models, both deterministic and stochastic, when many parameters should be estimated on the basis of unreliable data, especially for long-range estimates which are even more important for an outbreak with such dramatic consequences the whole world is experiencing. This inevitably results in a few more unreliability of the estimates. 

\begin{figure}
\centering
\includegraphics[width=1.00\textwidth]{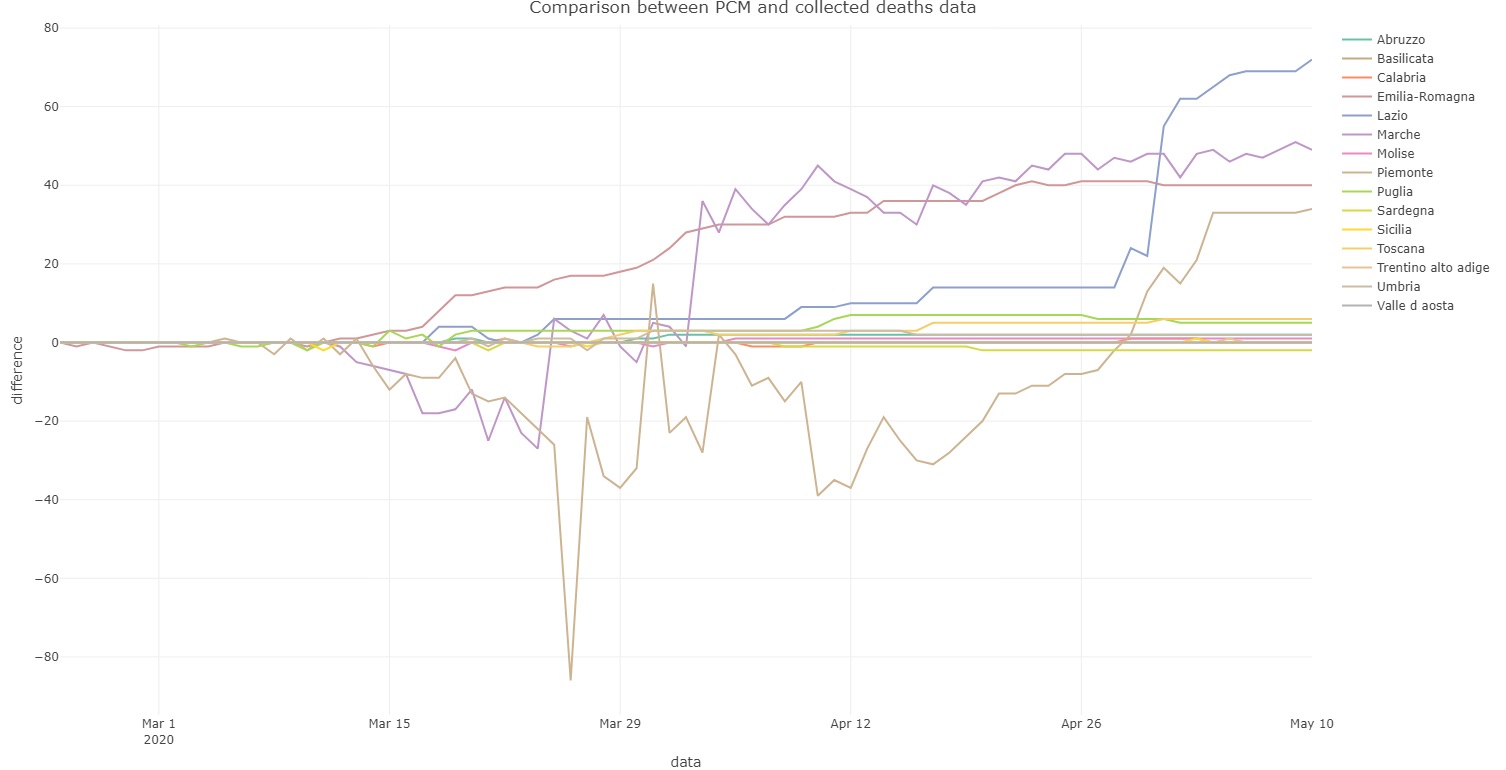}
\caption{Official and scraped data comparison}
\label{fig:data_comp}
\end{figure}

Figure \ref{fig:data_comp} reports on the differences between the sum of province COVID-19 cause specific deaths and the corresponding total regional deaths for each region for which we have all the province COVID-19 cause specific deaths from regional authorities in the February 24th - May 10th, 2020 time-span. For a few regions differences are often related to the delay between the time when data on deaths are published on press conferences and the time they are reported in official CPA data repository. This is an issue experienced also in other countries during the pandemic (see \cite{Seaman} for the number of deaths adjustments in UK). Figure \ref{fig:data_comp} clearly shows that for 5 regions out of 15 there are some discrepancies with respect to the official CPA data. 

\section{An adjusted time-dependent SIRD model}
\begin{figure}
\centering
\includegraphics[width=1.00\textwidth]{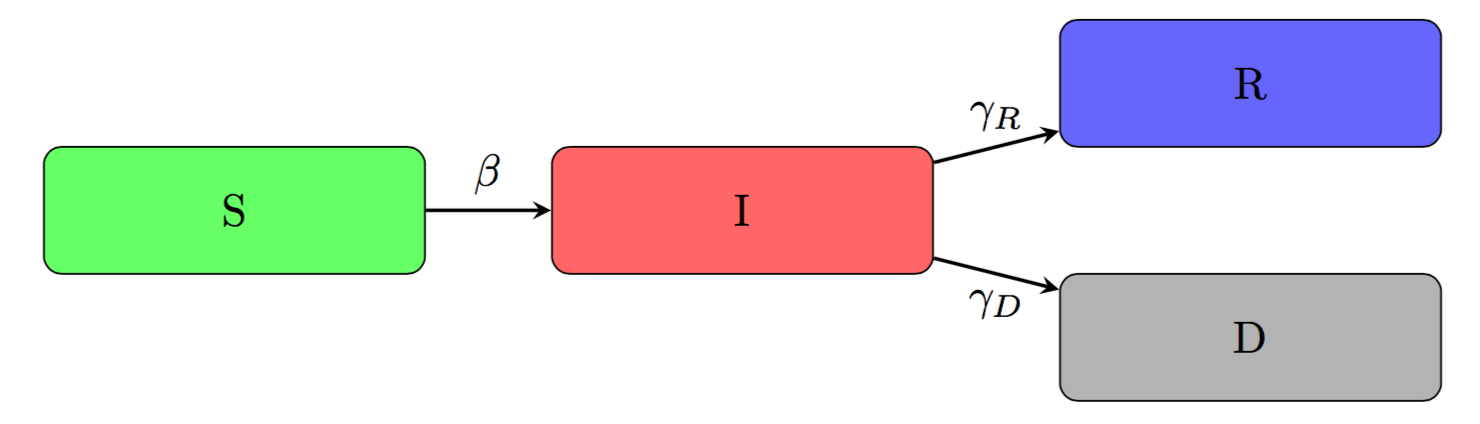}
\caption{SIRD model compartments and flows}
\label{fig:sird_comp}
\end{figure}

The SIRD model is a compartmental model used in epidemiology to design the spread of a disease \cite{Capasso,Kermack,Newman}. The model divides the population into four different groups: susceptible, infected, recovered, and deaths. This kind of design is appropriate when the disease of interest respects the following two assumptions: infected individuals must be infectious and recovered individuals receive longstanding immunity. The COVID-19 epidemic definitely respects the first assumption and some preliminary studies show that recovered individuals receive at least a short-term immunity. Other important assumptions concern the population: its size is considered fixed and individuals are identical to one another (i.e. demographic factors or health condition are not considered). In Figure \ref{fig:sird_comp} a schematic of the compartments and flows forming the model is shown.

The SIRD model is based on four variables $S(t)$, $I(t)$, $R(t)$, and $D(t)$, that are respectively the number of susceptible people, currently infected, total recovered and total deaths at time $t$. The size of the population is given by the sum of these four variables and it is represented by $n$. The parameters governing the model are the transmission rate, the recovery rate, and the mortality rate, that are respectively represented by $\beta$, $\gamma_R$, $\gamma_D$. Being rates, these parameters can also be seen respectively as the average time between effective contagious contacts ($\beta^{-1}$) and the average time before removal from the infectious class ($(\gamma_R+\gamma_D)^{-1}$).

Another important parameter that wholly describes the spread of an outbreak is the basic reproduction number, $R_0$, that is computed as the ratio between the transmission rate and the sum of recovery and mortality rate. $R_0$ represents the expected number of individuals that are directly infected by one infected individual, in a population where everyone is susceptible to infection. If $R_0$ is less than 1, the epidemic will eventually be controlled. If it is larger than 1, the transmission of the disease will increase in the population.

\begin{equation*}
    R_0 = \frac{\beta}{\gamma_R+\gamma_D}\label{r0}
\end{equation*}

Building on Chen et al. work \cite{Chenetal}, in this paper a time-dependent model is proposed in order to let the parameters be free to change over time. This kind of model is chosen because in Italy, as well as in various other countries facing the virus, containment measures have been adopted and incremented over time. In particular, a national lockdown was introduced in Italy on March 11th, 2020 and lasted  until May 4th, 2020. By allowing the parameters, and especially the effective transmission rate, to vary over time, the effect of control measures can be somewhat included in the model. On the other hand, recovery and mortality rate are likely to depend on the pressure under which hospitals and, in particular, intensive care units are in, which increases sharply at the beginning of an epidemic (i.e. when a high mortality rate is reported) and then relaxes after the health system capacity is enhanced.

The differential equations of the typical SIRD model are transformed into discrete time difference equations and result as follows:

\begin{eqnarray}
S(t+1)-S(t)&=&-\frac{\beta(t)S(t)I(t)}{n}\label{eqdiff1}\\
I(t+1)-I(t)&=&\left(\frac{\beta(t)S(t)}{n}-\gamma_R(t)-\gamma_D(t)\right)I(t)\label{eqdiff2}\\
R(t+1)-R(t)&=&\gamma_R(t) I(t)\label{eqdiff3}\\
D(t+1)-D(t)&=&\gamma_D(t)I(t)\label{eqdiff4}
\end{eqnarray}\\

From the records of the four variables of interest in a specific province, the evolution each parameters can be retrieved using the equations above. The time series are then used to predict the future values of transmission rate, recovery rate, and mortality rate. The method used in \cite{Chenetal} is a Finite Impulse Response (FIR) filter. The following equations describe the regression model used for each parameter and the cost function to be minimized, in order to find the optimal coefficients: 

\begin{eqnarray*}
\hat{Y}(t)=c_0+\sum_{j=1}^{J}c_jY(t-j)\label{ARmodel}\\
min_{c_j} \left(\sum_{t=J}^{T-2}(Y(t)-\hat{Y}(t))^2-\lambda\sum_{j=0}^{J}c^2_j\right)
\label{Ridge}
\end{eqnarray*}

where $Y(t)$ is one of the evolving parameters of the model, i.e. $\beta(t),\gamma_R(t),\gamma_D (t)$.
The FIR filter requires the choice of a single hyper-parameter (\(J\)) which represents the number of lag days to include in the regression. The cost function consists in a regularized least squares method in which the penalty (\(\lambda\)) is applied to the sum of squares of the regression coefficients (Ridge regression regularization, based on a $\ell_2$ norm \cite{Hoerl}). 

Once the models have been trained using historical data, future predictions can be done on the parameters and, therefore, estimates for the evolution of $S$, $I$, $R$, and $D$ can be computed using the SIRD model equations. 

\section{Adjusted training process}
The aim of our model is to make predictions about the evolution of the COVID-19 outbreak in Italy at the local level, particularly using historical data on each province. The model presented here differs from \cite{Chenetal} in that the hyperparameters are not fixed but optimized using different approaches. The model is composed by three different autoregressions based each on a SIRD model's parameter. Each of the regressions requires the choice of two hyperparameters: the number of lags ($J$) and the penalty for the regularization process ($\lambda$). As the transmission rate is the parameter that most depends on policy decisions and therefore the most likely parameter to change over time, the number of lags for the other parameters has been set equal to the optimal number of lags for $\beta(t)$, so as to have three homogeneous time series. The method that was employed to choose the optimal number of lags is the Akaike Information Criterion (AIC): the number of lags that minimizes the AIC is chosen within a given range. On the other hand, the regularization parameters were set free to change within a range for each of the parameters and cross-validation was employed to find their optimal values \cite{Stone}.

\begin{figure}
\centering
\includegraphics[width=1.00\textwidth]{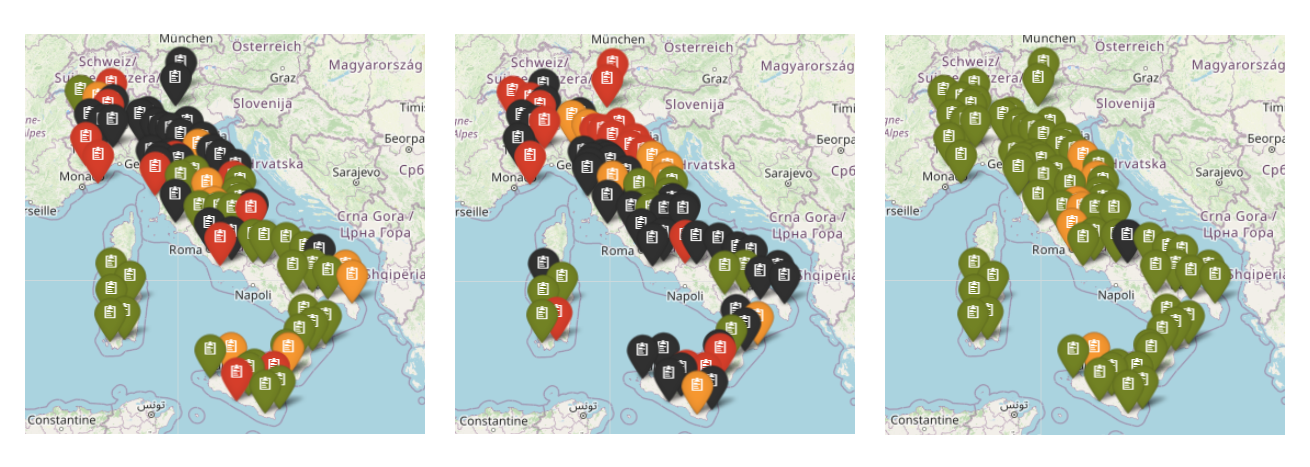}
\caption{Comparison of $R_0$ index values on March 10th (left), April 10th (center), and May 10th (right) for provinces where the number of deaths is available. Green pins are for $0 \leq R_0 < 0.5$, orange pins for $0.5 \leq R_0 < 1$, red pins for $1 \leq R_0 < 2$ and black pins for $R_0 \geq 2$}
\label{fig:r0_maps}
\end{figure}

However, the deterministic model described above might struggle in producing reliable estimates in contexts where the number of cases is very low and  there is considerable fluctuation or inconsistency in the data, as it happens in some provinces where the outbreak was not so intense (see Figure \ref{fig:r0_maps}, where the heterogeneity of the outbreak is clearly shown at provincial level, at least in the first-medium stages of its evolution), whilst it seems to give more robust results when data are aggregated at a higher level, as in the entire country's time series. This happens because the model is based on the smoothing of the sequential values of the variables that becomes less precise as the numbers became smaller. Therefore, different approaches for training were developed.

Firstly, a local-based training was considered; however this choice did not lead to robust results due to the issues highlighted above. Secondly, a training at the national level was suggested, as the lockdown was declared almost simultaneously in every Italian province. Nevertheless, the virus has spread irregularly in different geographical zones, with the southern areas reporting considerably fewer cases as well as a delay in the spread of COVID-19. A nation-based training process would have been highly influenced by the northern regions' critical situation and would have overestimated the development of the epidemic in the central and southern regions, far less affected from the virus.  Therefore, these two first methods were discarded.

Therefore, in a second attempt to use local data, the regional level was considered the most suitable option in order to balance the heterogeneity found in the epidemic evolution among different territories and the need for aggregated data.
For each province, the corresponding region is selected. In order to choose the regions whose situation most resembles the one of the selected region, a correlation measure between all of the regions was computed: this correlation was based on the new cases' daily time series divided by the regions' population and therefore was time-weighted so that the most recent days have a much bigger impact than the days at the beginning of the epidemic. The weighting process has been developed as follows:

\begin{eqnarray*}
    w_{t-d} &=& e^{30-d}\label{weight}
\end{eqnarray*}

where $t$ is the current day, $d=0,..., t-1$ is the number of days between the observation of interest and the current day, and $w_{t-d}$ is the weight assigned to that observation. The weights are then normalized so that their sum equals 1. The regions showing a high correlation with the region containing the province of interest (i.e. higher than the median value) are selected and their data is aggregated to compose the training set.

A final approach that was adopted exactly replicated the previous one but used provinces instead of regions as geographical units. This was not a viable solution at the beginning since data were only available for a few provinces and is still not accessible for some of the worst-hit, and therefore more informative, territories.

These two proposed training processes, called respectively \textit{regional aggregation training} and \textit{provincial aggregation training} provide both acceptable results and, therefore, are both considered suitable to make short-term predictions.
Model coefficients resulting from the training process are then applied to the local data in order to make predictions for each province with the same procedure followed previously: given the predictions about the future values of the parameters, the future values of the variables $S$,$I$,$R$, and $D$, are computed using the discrete time difference equations (\ref{eqdiff1})-(\ref{eqdiff4}).

\begin{figure}
\centering
\includegraphics[width=1.00\textwidth]{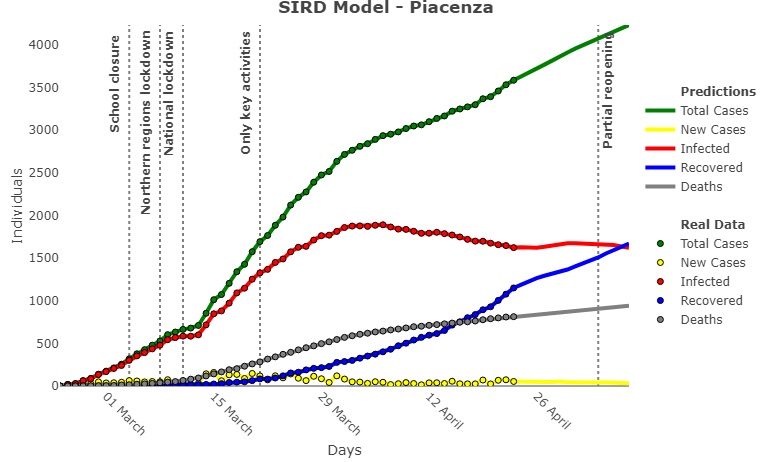}
\caption{Adjusted SIRD model for Piacenza province - April 23rd, 2020}
\label{fig:sird_piacenza}
\end{figure}

\begin{figure}
\centering
\includegraphics[width=1.00\textwidth]{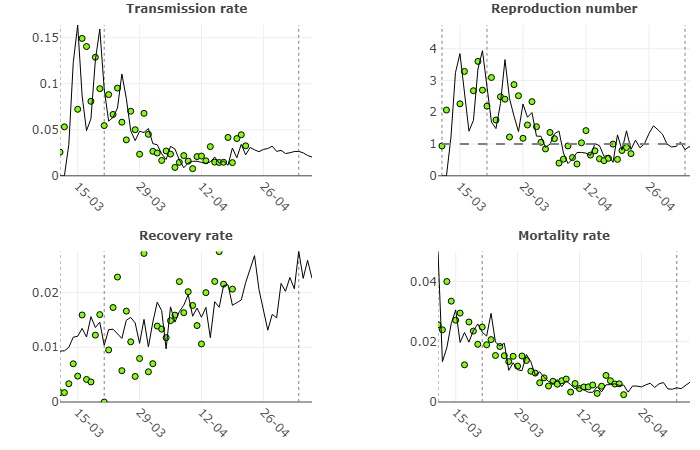}
\caption{Adjusted SIRD parameters predictions for Piacenza province - April 23rd, 2020}
\label{fig:sird_param}
\end{figure}

In Figures \ref{fig:sird_piacenza} and \ref{fig:sird_param}, an example on applying this model to the Piacenza province is shown. Piacenza is one of the provinces in northern Italy with the highest cumulative rate, since its main hospital is located very close to Codogno, the town with the first cluster of observed Covid-19 cases in Italy. In this case, the model is trained using regional aggregation. The vertical dashed lines represent the dates in which the containment measures were implemented.

Results for other provinces can be seen on a dashboard developed for this model, available at \url{https://ceeds.unimi.it/covid-19-in-italy/} (see also \cite{Ferrari} for a detailed description of this dashboard).

\section{Model accuracy evaluation}

\subsection{Comparison of training processes}
Due to the gradual decentralization of the health system competences to regional governments in the last years, the health policies implemented during the pandemic varied greatly throughout the country as well as over time. Moreover, the state of emergency during the pandemic in Italy has been detrimental to the data consistency: there has been numerous adjustments, corrections, and delays in the collection of the data, as well as changes in the definitions of the variables. The issue of data inconsistency is even more crucial to the sake of our model predictive power as it is focused on a quite disaggregated level. 

Our model's accuracy is evaluated using both the training approaches presented above. The Mean Absolute Percentage  Error \cite{Hyndman} (MAPE) is chosen as the accuracy metric to assess the error in model's predictions:

\begin{eqnarray*}
    MAPE_h &=& \frac{100}{PD} \sum_{p=1}^{P} \sum_{d=1}^{D} \left| \frac{Y_{hpd} - \hat{Y}_{hpd}} {Y_{hpd}}\right| \label{MAPE}
\end{eqnarray*}

where $h$ is the prediction horizon, $P$ is the number of provinces in the sample, $D$ is the number of days in the sample, $\hat{Y}_h$ is the resulting prediction and $Y_h$ is the real value for the variable. MAPE is a popular error measure used to asses the reliability of model prediction and is widely used in medical research (see, for example \cite{Aregay} or \cite{Prates}), even if it has some drawbacks \cite{McKenzie}.

The model is applied to all the provinces and within a fortnight period going from April 8th, 2020 to April 22nd, 2020 ($D=15$). For each day of the period, the model predicts the values of the variables using a time horizon from 1 to 15 days ahead in the future (i. e. $h=1,...,15$). The model's hyper-parameters are free to change (i.e. $\lambda$ and $J$ vary among provinces and among days). The model's predictions are then compared with the real values and the results are shown in Tables \ref{tab:regional} and \ref{tab:provincial}. In these tables, T stands for "Total cases".

\begin{table}[h]
 \small\sf\centering
  \caption{MAPE - Regional aggregation training}
    \begin{tabular}{|c|c|c|c|c|}
    \toprule
          Horizon  Days&  MAPE I &  MAPE R &  MAPE D &  MAPE T \\
          \midrule
           1 & 2.83 & 5.80 & 3.74 & 1.80 \\ 
  2 & 4.80 & 9.08 & 6.38 & 3.19 \\ 
  3 & 6.68 & 11.49 & 8.57 & 4.43 \\ 
  4 & 8.35 & 13.11 & 10.75 & 5.37 \\ 
  5 & 10.37 & 15.34 & 12.75 & 6.48 \\ 
  6 & 12.06 & 17.58 & 15.19 & 7.55 \\ 
  7 & 14.55 & 19.55 & 17.82 & 8.80 \\ 
  8 & 17.54 & 21.66 & 20.61 & 10.23 \\ 
  9 & 20.89 & 23.75 & 23.57 & 11.89 \\ 
  10 & 24.30 & 25.34 & 26.83 & 13.53 \\ 
  11 & 27.38 & 26.92 & 30.56 & 15.06 \\ 
  12 & 29.91 & 28.09 & 34.63 & 16.26 \\ 
  13 & 32.03 & 29.93 & 39.13 & 17.42 \\ 
  14 & 34.80 & 31.13 & 43.97 & 18.61 \\ 
  15 & 38.55 & 31.86 & 48.95 & 20.07 \\ 
     \bottomrule
     \end{tabular}
     \label{tab:regional}
 \end{table}
 
 \begin{table}[h]
 \small\sf\centering
  \caption{MAPE - Provincial aggregation training}
    \begin{tabular}{|c|c|c|c|c|}
    \toprule
         Horizon  Days&  MAPE I &  MAPE R &  MAPE D &  MAPE T \\
          \midrule
           1 & 2.88 & 5.70 & 3.34 & 1.67 \\ 
  2 & 5.16 & 9.15 & 5.91 & 2.94 \\ 
  3 & 7.25 & 11.58 & 8.19 & 4.17 \\ 
  4 & 9.35 & 13.83 & 10.66 & 5.26 \\ 
  5 & 11.68 & 16.55 & 13.00 & 6.44 \\ 
  6 & 14.29 & 19.34 & 15.52 & 7.72 \\ 
  7 & 17.31 & 21.95 & 18.19 & 9.29 \\ 
  8 & 20.94 & 24.17 & 21.02 & 11.11 \\ 
  9 & 24.86 & 26.07 & 24.38 & 13.16 \\ 
  10 & 29.07 & 28.14 & 28.19 & 15.19 \\ 
  11 & 32.77 & 29.42 & 32.05 & 17.14 \\ 
  12 & 37.06 & 30.98 & 36.24 & 19.35 \\ 
  13 & 41.79 & 32.21 & 40.52 & 21.85 \\ 
  14 & 48.33 & 35.73 & 45.29 & 24.68 \\ 
  15 & 80.42 & 88.28 & 50.18 & 27.55 \\ 
     \bottomrule
     \end{tabular}
     \label{tab:provincial}
 \end{table}
 
The MAPE is relatively low and acceptable for all of the variables in the short term. As in all time series, the error increases as the horizon for the prediction becomes larger: nevertheless, our model performance remains acceptable even in the longer-term, at least for the regional aggregation training approach. The estimate whose MAPE is the highest is for $R$ in both methods: note that $R$ has been estimated using regional data since reports about recovered at the provincial level are seldom published. The regional approach seems to perform slightly better than the provincial aggregation one, and its higher reliability is enhanced in the long term. The model is therefore considered the optimal training process and the analysis of its forecasting reliability is further developed.

\subsection{Error distribution} 
\label{sec:5_2}
We can look also at the distribution of the Mean Percentage Error (MPE) \cite{Hyndman} for each province to understand more about the prediction reliability of our model, given different time horizons, and highlight whether or not major bias issues arise: 

\begin{eqnarray*}
    MPE_{hp} &=& \frac{100}{D} \sum_{d=1}^{D}  \frac{Y_{hpd} - \hat{Y}_{hpd}} {Y_{hpd}} \label{MAPE2}
\end{eqnarray*}

\begin{figure}
\centering
\includegraphics[width=1.00\textwidth]{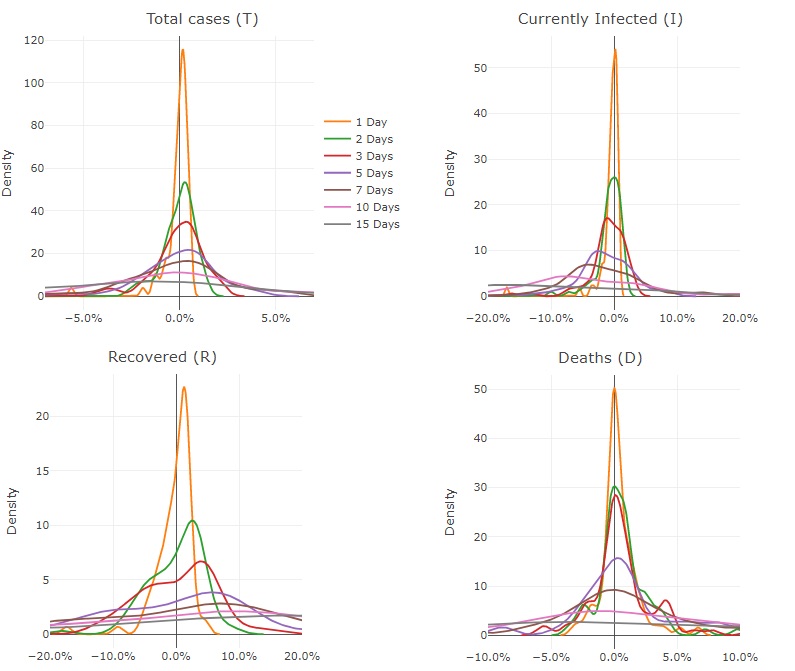}
\caption{MPE Distribution on different time horizons}
\label{fig:MPEDistr}
\end{figure}

The mean is computed all along the period so that each value represents the average error the model makes in the specific context of a province (Figure \ref{fig:MPEDistr}).

The distributions of the error for the total cases and the currently infected are similar. They both show mean values close to 0, low variances, and their errors are slightly negatively skewed. This means that the model tends to underestimate their values. However, the $I$ variable, since it depends not only on $\beta$ but also on the other two parameters, shows signs of higher uncertainty and its variance grows faster as the time horizon goes ahead in the future.

With regard to the recovered variable, its distribution is also negatively skewed meaning that the model tends to underestimate recovered as well, even though this could also be a consequence of the currently infected biased estimates.
Moreover, $R$ shows signs of high uncertainty, being the variable with the highest variance at low time horizons. As noted before, the values used for $R$ are not actually reported real data points but estimated computed using regional data. The distributions show that the lack of real data for this variable could be one of the major issues in the model results.

The deceased variable shows the best results in terms of accuracy in forecasting. Its mean is quite constant and close to 0 even as the time horizon grows. Its variance is still acceptable when the predictions are 15 days ahead in the future. Contrary to all the other variables, the MPE on $D$ show positive values for the mean as well as a positive skewness, meaning that, on average, the actual deaths are higher than the ones predicted. However, the skewness is not considered relevant and, in general, the model's performance for this variable is good. Detailed values for the MPE distribution analysis are reported in Table \ref{tab:provincial2}.

 \begin{table}[h]
 \small\sf\centering
  \caption{MPE Distribution Analysis}
    \begin{tabular}{|c|c|c|c|c|c|c|c|c|c|}
     \multicolumn{10}{c}{}\\
    \toprule
         & \multicolumn{3}{c|}{Mean} & \multicolumn{3}{|c|}{Standard Deviation} &   \multicolumn{3}{|c|}{Skewness} \\
          \midrule
           1 & -0.59 & -0.27 & 0.40 & 2.19 & 3.06 & 1.52 & -5.91 & -2.77 & 1.28 \\ 
  2 & -1.26 & -0.59 & 0.86 & 5.08 & 6.44 & 2.54 & -6.69 & -3.53 & 2.24 \\ 
  3 & -1.96 & -0.89 & 1.05 & 8.42 & 9.28 & 3.70 & -6.79 & -2.86 & 2.42 \\ 
  4 & -2.52 & -0.82 & 1.24 & 10.53 & 11.40 & 4.95 & -6.57 & -2.19 & 2.26 \\ 
  5 & -3.46 & -1.10 & 1.36 & 15.41 & 14.49 & 6.06 & -6.84 & -2.09 & 2.21 \\ 
  7 & -5.36 & -1.49 & 1.64 & 22.60 & 21.98 & 9.46 & -6.38 & -2.86 & 3.14 \\ 
  10 & -11.90 & -1.69 & 1.36 & 47.21 & 34.16 & 16.49 & -6.32 & -3.90 & 4.03 \\
  15 & -22.23 & -2.18 & 0.61 & 75.17 & 56.38 & 31.32 & -6.17 & -5.64 & 3.99 \\
       \bottomrule
     \end{tabular}
     \label{tab:provincial2}
 \end{table}
 
 The underestimation issue is probably due to an optimistic (i.e. lower) prediction  of the model about the future development of $\beta$, the transmission rate. One factor that could explain this issue and that is proved to be highly correlated with the number of new cases is the number of Covid-19 antigen tests conducted daily, which has been increasing since the beginning of the epidemic. This variable could not be included in the model due to data unavailability at the NUTS-3 level.
In addition to that, the policy concerning the reporting of each of the category might have changed over time and throughout the country, making it difficult for the model to adjust to those changes.

Although the model shows good results in the short-term, the mean values, as well as the variances, of the MPE tend to diverge from 0 as the time horizon widens. In order to choose a maximum reasonable time horizon for which the model's predictions can still be considered reasonable, the distribution of the error on currently infected is studied, since $I$ is the pivotal variable of the model. The MAPE is used again because the relevance is on the weight of the error. 

\begin{figure}
\centering
\includegraphics[width=0.80\textwidth]{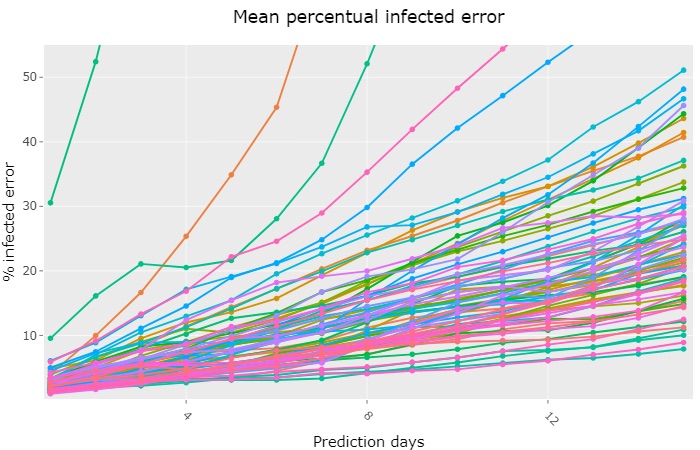}
\caption{Mean MAPE infected error by province using the regional aggregation training}
\label{fig:ABS_Inf_Err}
\end{figure}

Figure \ref{fig:ABS_Inf_Err} shows the evolution of the error with the growing days of prediction for each province using the regional aggregation training. The majority of the provinces shows low level of error even when the day of the prediction is 15. The plot in Figure \ref{fig:ABS_Inf_Err} also shows similar slopes in the growth of the MAPE for almost all of the provinces as the time horizon grows, with only few having unacceptable levels of MAPE displaying exponential growth.

\begin{figure}
\centering
\includegraphics[width=0.80\textwidth]{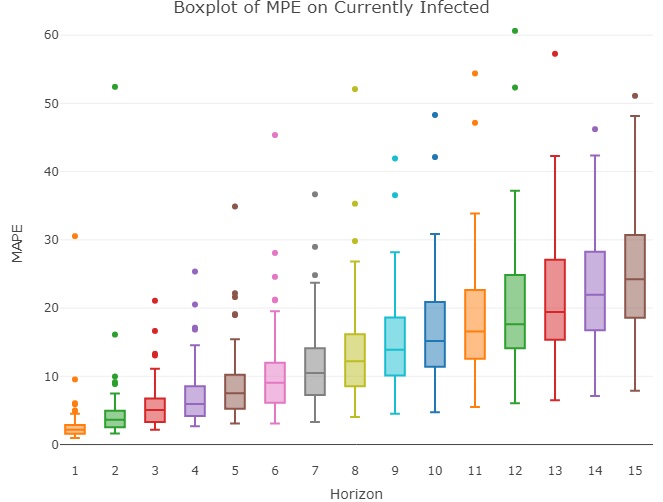}
\caption{Boxplots of MAPE on currently infected}
\label{fig:mape_boxplot}
\end{figure}

The boxplots in Figure \ref{fig:mape_boxplot} provide an insightful summary about the MAPE by province distribution for each day with respect to multiple horizons.
For a time horizon lower than 5 days, 75\% of the provinces are accurately predicted with percentage errors lower than 10\%. A time horizon equal to 6 days still shows an acceptable level of error with almost all of the provinces below the 20\% threshold. Even when the time horizon covers 13 days, half of the provinces show errors below 25\%, meaning that although the overall accuracy decreases, the model still performs pretty well on some provinces. The maximum reasonable time horizon is chosen using the value of the 3rd quartile metric and setting the upper bound to 20\%, so that at least 75\% of the provinces would show acceptable levels of prediction errors. Using this method, the resulting maximum time horizon is 9 days. We consider predictions within a 7 to 9-day period as being still accurate enough, while any consideration based on the model beyond this horizon may be misleading and should be carefully assessed.

\subsection{Clustering the provinces by one-week errors on $I$}
\begin{figure}
\centering
\includegraphics[width=0.80\textwidth]{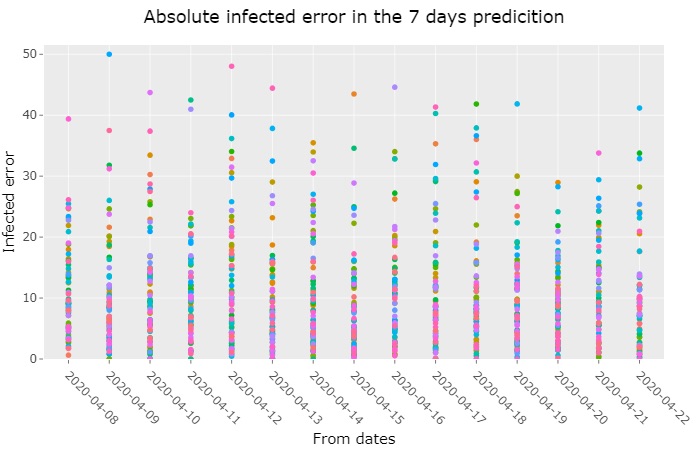}
\caption{MAPE on weekly predictions over time}
\label{fig:errors_weekly}
\end{figure}

On the basis of the conclusions of the error distribution analysis, weekly predictions are considered the optimal context of application for our model. Here, the MAPE on currently infected is used using a one-week horizon.
Figure \ref{fig:errors_weekly} shows how the one-week prediction error varies when we start to forecast the currently infected in different days. We can see that there are no alarming shifts in the model accuracy and that the large majority of the provinces are always below the 20\% threshold. Although this happens in the sample period of our choice, it is likely that the model accuracy has improved since the beginning of the epidemic, when most provinces reported very low numbers of cases, making the predictions harder.

\begin{figure}
\centering
\includegraphics[width=0.80\textwidth]{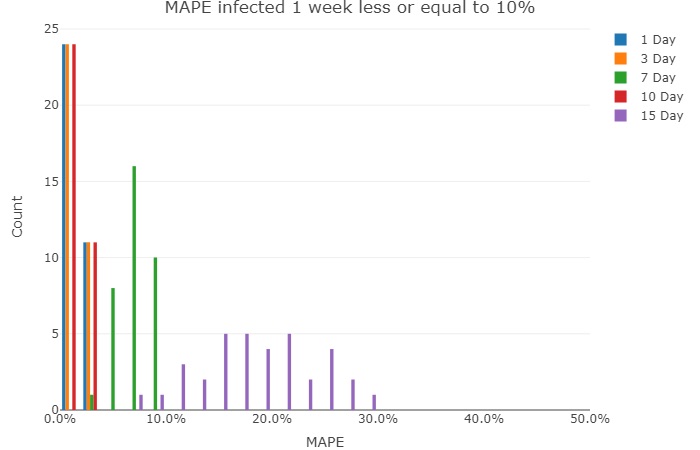}
\caption{MAPE infected for the provinces with prediction error on 1 week $\leq 10\%$}
\label{fig:errors_weekly1}
\end{figure}

\begin{figure}
\centering
\includegraphics[width=0.80\textwidth]{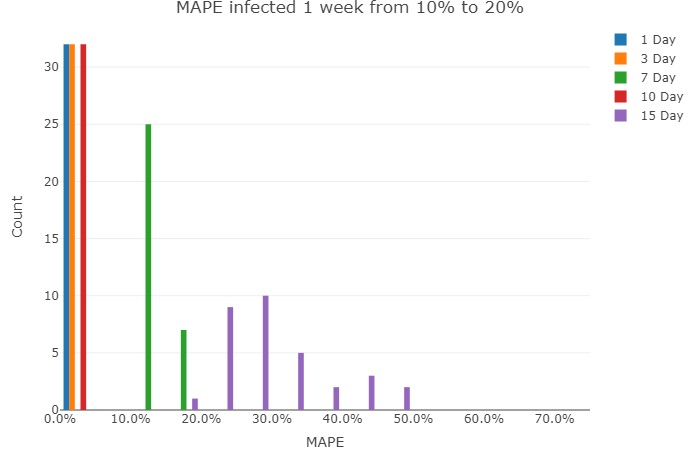}
\caption{MAPE infected for the provinces with prediction error on 1 week $> 10\%$ and $\leq 20\%$}
\label{fig:errors_weekly2}
\end{figure}

\begin{figure}
\centering
\includegraphics[width=0.80\textwidth]{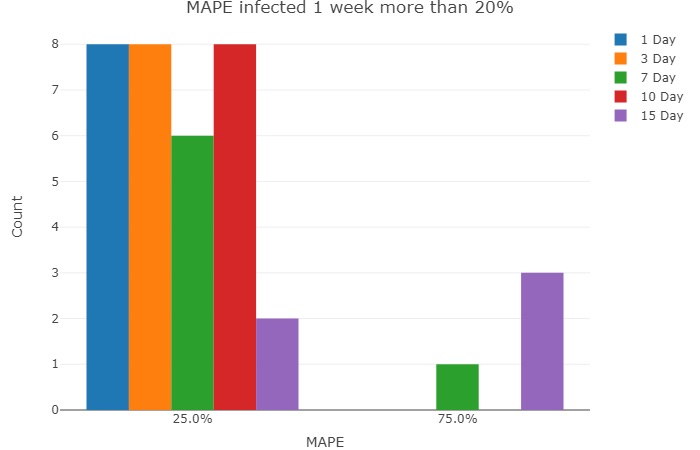}
\caption{MAPE infected for the provinces with prediction error on 1 week $ > 20\%$}
\label{fig:errors_weekly3}
\end{figure}

We now present the results on the average weekly MAPE by province for the currently infected $I$ (the variable used to select the correlated time series) using regional correlation. Almost half of the provinces show an average MAPE below 10\% and only 8 perform badly on average, with a weekly MAPE above 20\% (Figures \ref{fig:errors_weekly1}, \ref{fig:errors_weekly2} and \ref{fig:errors_weekly3}). 
There are different reasons why the model is unable to accurately forecast the development of the disease in these provinces. For example, some of them, such as Oristano and Matera, are among the least-hit provinces in Italy. On the other hand, there are some provinces like Imperia that show quite inconsistent time series on the official reported total cases, with frequent later adjustments. The following is the complete list of provinces with a MAPE less than 10\%, between 10\% and 20\% and more than 20\%.
\begin{itemize}
    \item MAPE$_I$ $\leq$ 10\% (Figure \ref{fig:errors_weekly1}): 
    Agrigento, Alessandria, Bari, Biella, Bologna, Brindisi, Caltanissetta, Campobasso, Catanzaro, Crotone, Enna, Firenze, Forli-Cesena, Frosinone, L'Aquila, Latina, Lecce, Massa Carrara, Modena, Palermo, Piacenza, Pistoia, Reggio nell'Emilia, Rimini, Roma, Sassari, Siena, Sud Sardegna, Taranto, Teramo, Torino, Trapani, Vercelli, Vibo Valentia, Viterbo (n = 35; mean = 6.95; sd = 1.49; skewness = -0.66; kurtosis = 2.78).
    \item 10\% $<$ MAPE$_I$  $ \leq$ 20\% (Figure \ref{fig:errors_weekly2}): Ancona, Ascoli Piceno, Asti, Barletta-Andria-Trani, Cagliari, Catania, Chieti, Cosenza, Cremona, Cuneo, Fermo, Ferrara, Foggia, Grosseto, Livorno, Lucca, Macerata, Messina, Novara, Nuoro, Parma, Pesaro e Urbino, Pescara, Pisa, Potenza, Prato, Ragusa, Ravenna, Reggio Calabria, Rieti, Siracusa, Verbano-Cusio-Ossola (n = 32; mean = 13.44; sd = 2.68; skewness = 0.92; kurtosis = 3.14).
    \item MAPE$_I$ $>$ 20\% (Figure \ref{fig:errors_weekly3}): Aosta, Arezzo, Imperia, Isernia, Matera, Oristano, Perugia, Terni (n = 8; mean = 52.18; sd = 59.17; skewness = 2.01; kurtosis = 5.41).
\end{itemize}

\subsection{Bootstrap prediction intervals}
\begin{figure}
\centering
\includegraphics[width=0.80\textwidth]{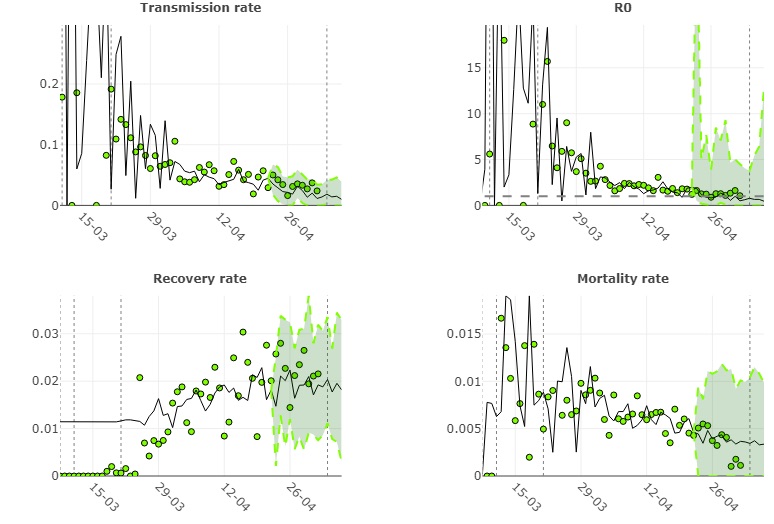}
\caption{Parameters' prediction intervals with $\alpha$ = 10 \% - Torino province}
\label{fig:intervals_param}
\end{figure}

Another way to assess the model reliability and the confidence level at different time horizons is by estimating prediction intervals for the parameters, and in turn, for the variables.
In order to build prediction intervals for the model's parameters, a forward bootstrap algorithm using fitted residuals was used \cite{Politis}. A zero lower bound was imposed for all the parameters so as to avoid results that contradict the compartmental logic of the SIRD model. The number of lags in each regression is kept constant during the bootstrap re-training repetitions. The same procedure is applied to the three regressions of the model, respectively used for $\beta$, $\gamma_R$, and $\gamma_D$. The lower ($low$) and upper ($up$) bounds for the basic reproduction number are computed using Equation (\ref{R0_lower}). However, when all the three parameters have a zero lower bound, the upper bound of $R_0$ cannot be computed since the denominator will be equal to 0. Bootstrap intervals for each parameter are shown in Figure \ref{fig:intervals_param}.

\begin{eqnarray}
    R_0(t)_{low/up}&=&\frac{\beta(t)_{low/up}}{\gamma_R(t)_{up/low}+\gamma_D(t)_{up/low}}\label{R0_lower}
\end{eqnarray}

Each of the model's variables depends on all of the parameters so that the obtained error range for a parameter must be combined with the other two parameters' intervals: thus, the intervals for the variables are subject to the uncertainty of three different parameters and can be composed in different ways.

Combining lower and upper bounds of the parameters can be misleading since the variables of the SIRD model develop in different directions and, especially, each variable depends on the past value of $I(t)$ which in turn depends on $S(t-1)$, $R(t-1)$, and $D(t-1)$. Nevertheless, combinations of interest can be used to describe the epidemic development in particular scenarios.
The method proposed here is to use the prediction interval for the parameter of interest and use point estimates for the other two parameters. Thus, the effects of the variability of the parameter can be easily displayed on each variable. According to this method, when the parameter of interest is $\beta$, the following equations are used to compute the prediction intervals for each variable:
\begin{eqnarray*}
S(t+1)_{low/up}&=&S(t)_{low/up}\left(1-\frac{\beta(t)_{up/low}I(t)_{up/low}}{n}\right)\label{lowerS}\\
R(t+1)_{low/up}&=&R(t)_{low/up}+\gamma_D(t)I(t)_{low/up}\label{lowerR}\\
D(t+1)_{low/up}&=&D(t)_{low/up}+\gamma_R(t)I(t)_{low/up}\label{lowerD}\\
I(t+1)_{low/up}&=&n-S(t+1)_{up/low}-R(t+1)_{low/up}-D(t+1)_{low/up}\label{lowerI}
\end{eqnarray*}\\

\begin{figure}
\centering
\includegraphics[width=0.80\textwidth]{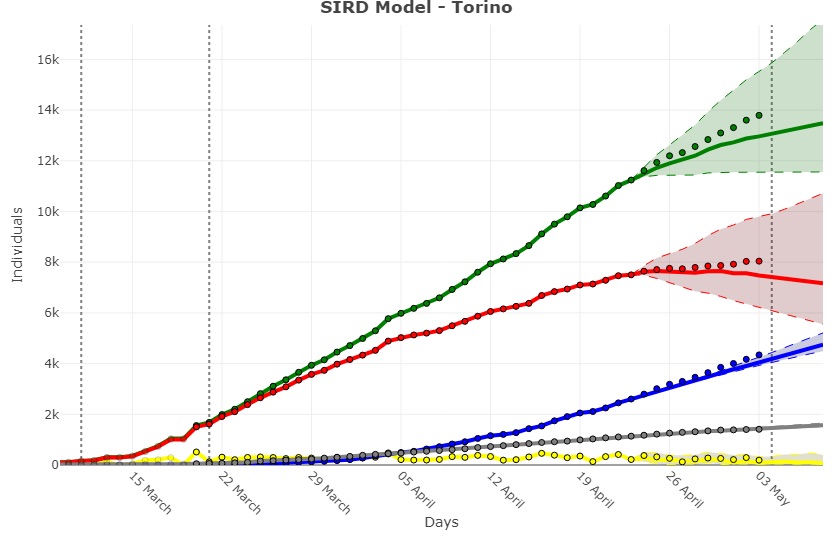}
\caption{Variables' prediction intervals based on $\beta$ interval - Torino province}
\label{fig:intervals_var}
\end{figure}

Bootstrap intervals for each variable are shown in Figure \ref{fig:intervals_var}.

\section{Conclusion and further developments}
The model presented in this paper can be applied at the NUTS-3 region level in Italy in order to predict the future development of the epidemic in the specific provincial context. Although two main training approaches were presented, the \textit{regional aggregation training} has proved more accurate and it is therefore recommended.

The main issue found during the building of our model was the lack of detailed and consistent data about the epidemic at the provincial level: if more variables had been available, the model could have been extended to include other compartments, such as the hospitalized cases or the number of tested individuals. The recovered individuals data was also not available and, therefore, an estimate using regional data was necessary. Finally, the state of emergency in which the data was collected is likely to have affected its quality and its consistency. 
The sum of these issues has definitely been detrimental to the model predictive power.
Nevertheless, the model seems to perform relatively well in the short-term up until a 9-day horizon and with the variable we were able to scrap from regional authorities and local newspaper websites.

Predictions about the peak day or the "0 new cases" day can also be done but they are likely to be far less accurate than the ones made with a wider geographical perspective, such as the national level, given the local context and the low numbers our model faces.
Nonetheless, the model presented can be useful to gain a general understanding about the epidemic development in the short-term at the local level. Particularly, it could be implemented to monitor and signal the provinces at greater risk in the near future.

Along with the forecasting of the SIRD variables, the model offers other insights about the epidemic development. The computation of the parameters provide three informative time series about the evolution of the disease in each specific context. 
For example, the analysis of the transmission rate $\beta(t)$ evolution within a province or the comparison among multiple ones can be extremely useful to gain additional information about which and how containment measures have been effective. 
On the other hand, the values of recovery rate and mortality rate can highlight issues and strengths of the health system across the country in small areas. 

Future work will be about considering an extended stochastic version of the SIRD model presented in this paper (for example, building on the work by Zimmer et al. \cite{Zimmer}), the Bayesian framework and multiple-source models as in \cite{DeAngelis}.

\bibliographystyle{SageV}

\end{document}